\documentclass[11pt,english,notitlepage]{article}
\usepackage{times}
\usepackage[T1]{fontenc}
\usepackage[latin1]{inputenc}
\usepackage{geometry}
\usepackage{cite}
\usepackage{amsmath, amssymb, psfrag, graphicx, color, tabls,  babel}
\usepackage{multicol}

\newcommand{\beq}{\begin{eqnarray}}
\newcommand{\eeq}{\end{eqnarray}}
\newcommand{\beqn}{\begin{equation}}
\newcommand{\eeqn}{\end{equation}}

\newcommand{\vc}[1]{V\!(\text{C}|#1)}
\newcommand{\vd}[1]{V\!(\text{D}|#1)}
\newcommand{\id}[1]{\mathbf{1}_{(#1)}}

\begin{document}

\centerline{\textbf{\LARGE{}The effect of finite population size on the evolutionary}}\vspace{.5ex}
\centerline{\textbf{\LARGE{}dynamics in multi-person Prisoner's Dilemma}} 
\vspace{2ex}
\centerline{\large Anders Eriksson$^\ast$ and Kristian Lindgren$^\dagger$}
\vspace{1ex}
\centerline{\small Division of Physical Resource Theory -- Complex Systems, Dept. of Energy and Environment} \centerline{\small Chalmers  University of Technology, SE-41296 Sweden}
\centerline{\small Email: $^\ast$anders.eriksson@chalmers.se $^\dagger$kristian.lindgren@fy.chalmers.se}

\begin{abstract}
\noindent
We study the influence of stochastic effects due to finite population size in the evolutionary dynamics of populations interacting in the multi-person Prisoner's Dilemma game. This paper is an extension of the investigation presented in a recent paper [Eriksson and Lindgren (2005), J. Theor. Biol. 232(3), 399].
One of the main results of the previous study is that there are modes of dynamic behaviour, such as limit cycles and fixed points, that are maintained due to a non-zero mutation level, resulting in a significantly higher level of cooperation than was reported in earlier studies.
In the present study, we investigate two mechanisms in the evolutionary dynamics for finite populations: (i) a stochastic model of the mutation process, and (ii) a stochastic model of the selection process. The most evident effect comes from the second extension, where we find that a previously stable limit cycle is replaced by a trajectory that to a large extent is close to a fixed point that is stable in the deterministic model. The effect is strong even when population size is as large as $10^{4}$. The effect of the first mechanism is less pronounced, and an argument for this difference is given.

\end{abstract}

\section{Introduction}
\label{sec:Introduction}

When several individuals interact in a group to produce a good or to receive
a benefit, the best strategy for a player depends on the details of the game as well as 
on the strategies of the others. For the group as a whole, and often for each individual in the long-term
perspective, it would be best if cooperation could be established.
In social and natural systems, there are, though, numerous
examples of situations where so-called free riders or defectors
take advantage of others cooperating for a common good
\cite{hardin68,maynard-smith82,sugden86}.
A game-theoretic approach for the study of cooperation can be
based on the Prisoner's Dilemma game
\cite{flood58,rapoport_chammah65} -- a situation that captures
the temptation to act in a selfish way to gain a higher own reward
instead of sharing a reward by cooperating. In the game, the
players independently choose an action, either to defect or to
cooperate.

In the two-person game, the scores are $R$ (reward) for mutual
cooperation, $T$ (temptation score) for defection against a
cooperator, $S$ (sucker's payoff) for cooperation against a
defector, and $P$ (punishment) for mutual defection, with the
inequalities $S < P < R < T$ and (usually) $R > (T+S)/2$. We use
fixed values on $R$ and $S$ in this study, $R = 1$ and $S = 0$, while
$0 < P < 1 < T < 2$; in the population dynamics we use there are
only three independent parameters, the third one being a growth
constant.
From theoretic and simulation studies of two-person Prisoner's
Dilemma game, it is known under which circumstances repeated
interactions may allow for a cooperative population to be
established that can resist invasion by non-cooperative mutants
(see, e.g., \cite{rapoport_chammah65, axelrod_hamilton81,
molander85, axelrod87, miller96, lindgren92, nowak_sigmund92}).

In the $n$-person Prisoner's Dilemma game, $n$ players
simultaneously choose whether to cooperate or to
defect. In the literature, there are several evolutionary models
based on the $n$-person Prisoner's Dilemma using various strategy
sets and pairing mechanisms, e.g., where the players are
distributed in space (see,
e.g., \cite{matsuo85, adachi_matsuo91, adachi_matsuo92,
albin92, hauert_schuster97, akimov_soutchanski97,
matsushima_ikegami98, lindgren_johansson01}).
In a recent paper \cite{eriksson_lindgren05}, we revisited the classic $n$-person Prisoner's
Dilemma. Following Boyd and Richersson\cite{boyd_richerson88} and
Molander \cite{molander92}, the behaviours of the participants were
modelled by simple reactive strategies. These authors analyse the
stability of stationary populations in the limit where mutations
are infrequent: a mutation either is driven to extinction by the
selective pressure from the resident population, or leads to a new
resident population. The general conclusion from their studies is
that cooperation is difficult to obtain when extending the group
size beyond the two persons in the original Prisoner's Dilemma
game.
In this limit, we have found \cite{eriksson_lindgren05} that for some values of the
payoff parameters, the rate of convergence to the evolutionarily
stable populations is so low that the assumption that mutations in
the population are infrequent on that time scale is unreasonable.
Furthermore, the problem is compounded as the group size is
increased. In order to address this issue, we derived a
deterministic approximation of the evolutionary dynamics with
explicit, stochastic mutation processes, valid when the population
size is large.

The question is: does the deterministic replicator dynamics introduced in \cite{eriksson_lindgren05} accurately describe the time evolution of the frequencies of strategies in a finite population?
In the present study, we investigate two mechanisms in the evolutionary dynamics for finite populations: (i) a stochastic model of the mutation process, and (ii) a stochastic model of the selection process. 

\section{The $n$-person Prisoner's Dilemma game}
\label{sec:The multi-person Prisoner's Dilemma game}

In the $n$-person Prisoner's Dilemma game each player
interacts with $n-1$ other players. Depending on the number $k$ of
others cooperating, a player receives the score $\vc{k}$ when
cooperating and the higher score $\vd{k}$ when defecting.
In order for the model to be well-behaved,  the score functions must obey two constraints: first, the
scores must increase with an increasing number of cooperators. Second,
the sum of the scores given to all players should increase if one player
switches from defection to cooperation (see, e.g.,
\cite{boyd_richerson88}). 
In this paper we shall assume that the scores $V$ can be
calculated as a linear combination of the scores against the other
players in $n-1$ ordinary two-player Prisoner's Dilemma games:
\beq
   \vc{k} &=& \frac{k}{n-1} \text{ and } \vd{k} = T\,\frac{k}{n-1} + P\,\frac{n - k - 1}{n-1} \, ,
\eeq
where we have divided by $n-1$ in order to make it easier to
compare results from different group sizes. The parameters $P$ and
$T$ obey $0 < P < 1 < T < 2$. Note that this is still an
$n$-person game since the same action is performed simultaneously
in all games. It is straight-forward to extend this model to arbitrary score functions $\vc{k}$ and $\vd{k}$ (provided the above constraints are fulfilled). The qualitative conclusions of this paper, however, are not expected to depend on the choice of score functions.

We focus on the set of trigger strategies \cite{schelling78} as
the strategy space for the evolution, which was also considered
by, e.g., Boyd and Richersson \cite{boyd_richerson88} and Molander \cite{molander92}. 
Despite their simplicity, trigger strategies capture many important aspects of
the many-person game, and allow for straight-forward evaluation of
the expected score for a player in a group randomly generated from
a given population. 
A trigger strategy $s_k$ is characterised by the degree of
cooperation that it requires in order to continue to cooperate: a
player with trigger strategy $s_k$ cooperates if at least $k$
other players cooperate. In a game with $n$ participants, $k$ is
in the range $0,\,\dots,\,n$. The strategy $s_0$ is an
unconditional cooperator and $s_n$ is an unconditional defector.
Each player decides whether to cooperate or to defect based on the
actions of the other players. In the first round after the
formation of a group, all players are assumed to cooperate, with
the exception of unconditional defectors. Then the players that
are unhappy with the number of cooperators switch to defection.
This may cause other players to change their action, and this is
iterated until a stable configuration has been reached. Note that
the number of cooperators may only decrease or be stable, and that
this procedure converges to the stable configuration with the
maximum number of cooperators. In a repeated game without noise,
this implies that a group of players with different trigger levels
reaches a certain degree of cooperation, some players may be
satisfied and cooperate while the others defect. We use the scores for the players in this 
equilibrium state to determine the selection, described in the next section.

\section{Evolutionary dynamics}
\label{sec:infinitesimal alpha}

Consider a population of $N$ individuals. From one generation to
the next, a fraction $\delta$ of the population is replaced using
fitness proportional selection, where the fitness of an individual
is proportional to the number of offspring surviving to
reproductive age. Throughout this paper, $\delta = 0.1$. If small
enough, the value of $\delta$ does not influence the structure of
the evolving population, but determines the evolutionary time
scale. Assuming that the population size is large and constant,
the evolutionary dynamics takes the form of
\beq\label{eq:replicator_dynamics_simple}
   x'_l &=& x_l + \delta \left( \frac{f_l}{\bar{f}} - 1 \right) x_l \, ,
\eeq
where $x_l$ is the fraction of players in the population with
trigger level $l$, $x'_l$ is the value of $x_l$ in the next
generation, $f_l$ is the expected fitness for a player with
trigger level $l$, and $\bar{f} = \sum_{l=0}^n x_l  f_l$ is
the average fitness in the population. The expected fitness $f_l$ for a player with trigger level $l$ is
the expected score of the player in a randomly formed group:
\beq\label{eq:payoff}
    f_l &=& \sum\limits_{i_1,\, \dots,\, i_{n-1} =\, 0}^n
           x_{i_1} \cdots\ x_{i_{n-1}} \ S(l, i_1, \ldots ,i_{n-1})
\eeq
where $S(l, i_1, \ldots ,i_{n-1})$ is the score of a player with
strategy $s_l$ in a game with $n-1$ other players, using
strategies $s_{i_1},\, \ldots,\, s_{i_{n-1}}$ respectively. 

Molander \cite{molander92} has analysed this model -- with general
score functions $\vc{k}$ and $\vd{k}$ -- under the assumption that a
mutation will either lead to a new resident population, or that
the evolutionary dynamics (\ref{eq:replicator_dynamics_simple}) will bring the population back to the original population before the next mutation occurs. Molander has shown that in each
interval $P \in (\frac{k-1}{n-1}, \frac{k}{n-1})$, where $k \in
\{1,\dots,n-2\}$, there is either a mixture of strategies $s_k$
and $s_n$, which is evolutionarily stable, or there is a mixture
of strategies $s_0,\ldots,s_{n-1}$ (all cooperating), that resists
invasion by strategy $s_n$, but which is not evolutionarily
stable. Finally, there is no other asymptotically stable
population in that interval. In the interval $P \in
(\frac{n-2}{n-1},1)$, the purely cooperative equilibrium mixture
is the only possible asymptotically stable population.

\begin{figure}

\centerline{\includegraphics{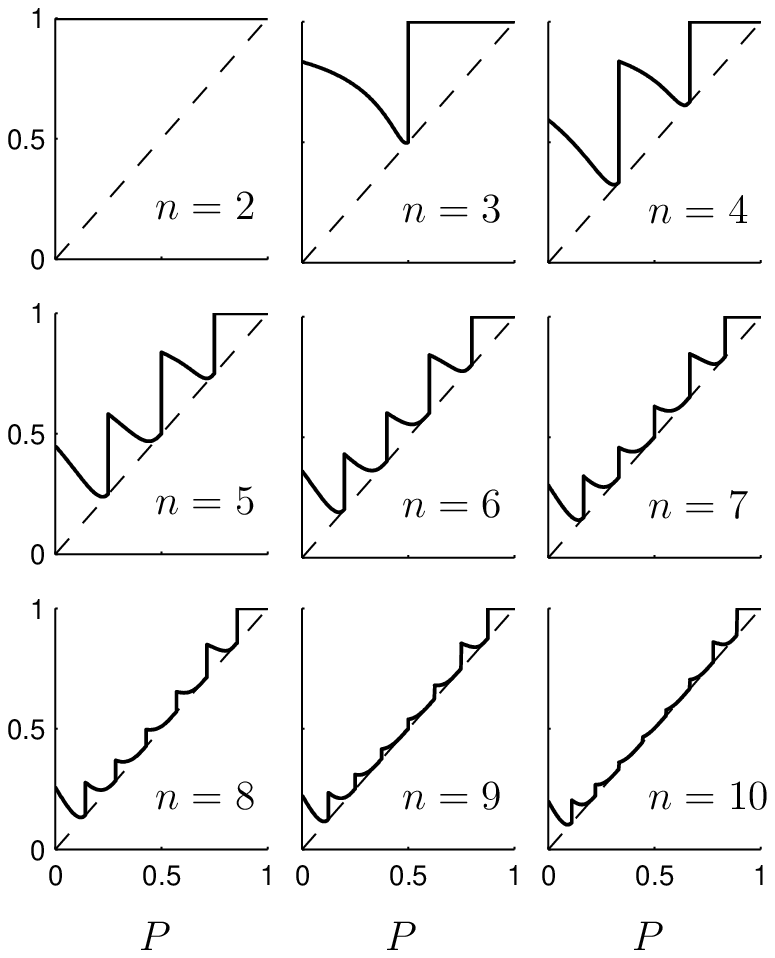}}

\caption{\label{fig:analytical_payoffs} The payoff at equilibrium
as a function of $P$ (thick line), for  $T = 1.2$ and group size
$n \in \{2,\dots,10\}$. Also shown is the payoff $P$ (dashed line)
for a population of pure all-defect. (Adapted from \cite{eriksson_lindgren05}.)
}
\end{figure}

Consider a population with groups of size $n$ consists of a
mixture of strategies $s_k$ and $s_n$, in fractions $x$ and $1-x$,
respectively. Since, in this population, strategy $s_k$ cooperates
if and only if there are at least $k$ other players with the same
strategy in the group, and since strategy $s_n$ always defects,
direct evaluation of (\ref{eq:payoff}) gives
\beq
    f_k(x) &=& \sum_{i=1}^k P \binom{n-1}{i-1} x^{i-1} (1-x)^{n-i} +
    \sum_{i=k+1}^n \frac{i-1}{n-1} \binom{n-1}{i-1} x^{i-1} (1-x)^{n-i}
\eeq
for strategy $s_k$ and
\beq
    f_n(x) &=& \sum_{i=0}^k P \binom{n-1}{i} x^i (1-x)^{n-i-1} + \nonumber\\
    		&& +\ \sum_{i=k+1}^{n-1} \left[P + (T - P) \frac{i}{n-1} \right] \binom{n-1}{i} x^i (1-x)^{n-i-1},
\eeq
for strategy $s_n$. We find the equilibrium by setting $f_k(x) =
f_n(x)$ and then solving numerically for $x$, with the requirement $0 < x <
1$. Existence and uniqueness of this equilibrium is guaranteed by
the result of Molander \cite{molander92}. 
In Fig.~\ref{fig:analytical_payoffs} we show how the equilibrium
fitness depends on $P$ for $T = 1.2$, for group size $n \in
\{2,\dots,10\}$. 
The fitness at the asymptotically stable population approaches $f = P$ as $n$ increases, indicating
a decreasing degree of cooperation with increasing $n$.  From the existence and
uniqueness of the asymptotically stable populations of this form,
and from a Taylor expansion of $f_k$ and $f_n$ to the $k+1$th order, follows that $x_k \propto 1/n$ at the asymptotically stable
population, for $T > 1$. The decrease in cooperation in the evolutionarily stable populations is what has led earlier authors \cite{boyd_richerson88, molander92} to conclude that cooperation in large or event modestly sized groups is evolutionarily unfavoured.

The number of terms in direct evaluation of the expected fitness from (\ref{eq:payoff}) grows very rapidly with $n$. Hence, in order to simulate the evolutionary dynamics (\ref{eq:replicator_dynamics_simple}) for general population compositions, it is necessary to have an efficient method for evaluating the expected fitness of each strategy. Using the probability $P^l_i$ that the number of cooperating players equals $i$, in a group with one player using strategy $s_l$ and the $n-1$ other players chosen randomly from the population, we may express the expected fitness $f_l$ as
\beq\label{eq:payoff_l_final}
    f_l &=& \sum_{i=0}^l P^l_i \, \vd{i} + \sum_{i=l+1}^n P^l_i \, \vc{i-1},
\eeq
since strategy $s_l$ cooperates if $i > l$. Thus, efficient calculation of the $P^l_i$ allows for the study of the evolutionary dynamics in groups of more than a few players. Since $P^l_i$ only depends on the distribution of trigger levels in the population, this method may be applied for any payoff functions $\vc{i}$ and $\vd{i}$. In the previous paper \cite{eriksson_lindgren05} we derived the following formula for $P^l_i$:
\beq\label{eq:P forumla}
    P^l_i &=& \left\{
 \begin{array}{ll}
        0 & \text{ whenever } i = l, \\
        D^{l,0}_{n-1} & \text{ when } i = 0, \\
        \left( x_0 + \ldots + x_{n-1} \right)^{n-1} & \text{ when } i = n, \\
        \binom{n-1}{i - \id{l < i}}\ 
        \left( x_0 + \ldots + x_{i-1} \right)^{i - \id{l < i}}
        D^{l,i}_{n - 1 - i + \id{l \,<\, i}}
        & \text{ otherwise, }
    \end{array}
    \right.
\eeq
where $\id{l \,<\, i}$ is one if $l < i$ and zero otherwise, and $D^{l,i}_m$ is given by the recursive formula
\beq\label{eq:D formula}
    D^{l,i}_m &=& \left\{
 \begin{array}{ll}
    x_n^m & \text{ when } i = n - 1 \\
    \sum\limits_{j = 0}^{M} \binom{m}{j}\, x_{i+1}^j\, D^{l,i+1}_{m - j} & \text{ otherwise}
    \end{array}
    \right.
\eeq
where $M = m + i + 2 - \id{l \,<\, i} - n$. Note that whenever $l < i$ and $l' < i$, $P^l_i = P^{l'}_i$, so if $l < i$ then $P^l_i  =  P^0_i$. Using tables to store evaluated values of $D^{l,i}_m$, it is possible to evaluate the values of $P^l_i$ for all $l$ and $i$ in $\sim n^4$ operations. 

\section{Selection and mutation processes in finite populations}

\begin{figure}
\centerline{
	\psfrag{x1}[][t]{$P$}
	\psfrag{y1}[][]{$\bar{f}$}
	\psfrag{atext2}{}
	\psfrag{n = 6}[][]{\small$n = 6$}
	\includegraphics[clip,height=5cm]{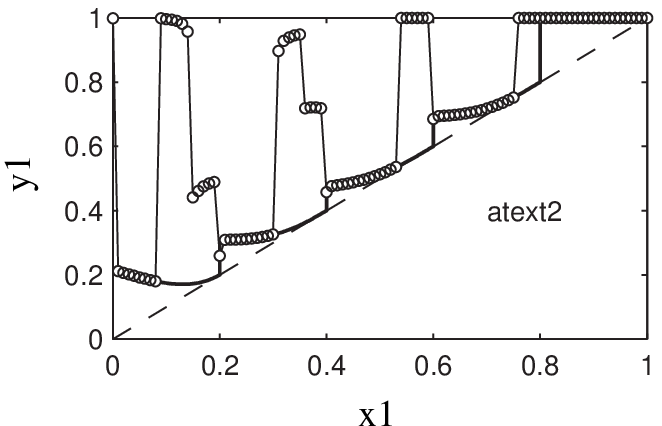}
	\hspace{-7.5cm}\raisebox{4.5cm}{\textbf{a}}\hspace{7.5cm}
	\raisebox{6pt}{
		\psfrag{xlabel}[t][]{$\mu\,t$}
		\psfrag{ylabel}[b][]{$x_i(t)$}
		\includegraphics[clip,height=4.8cm]{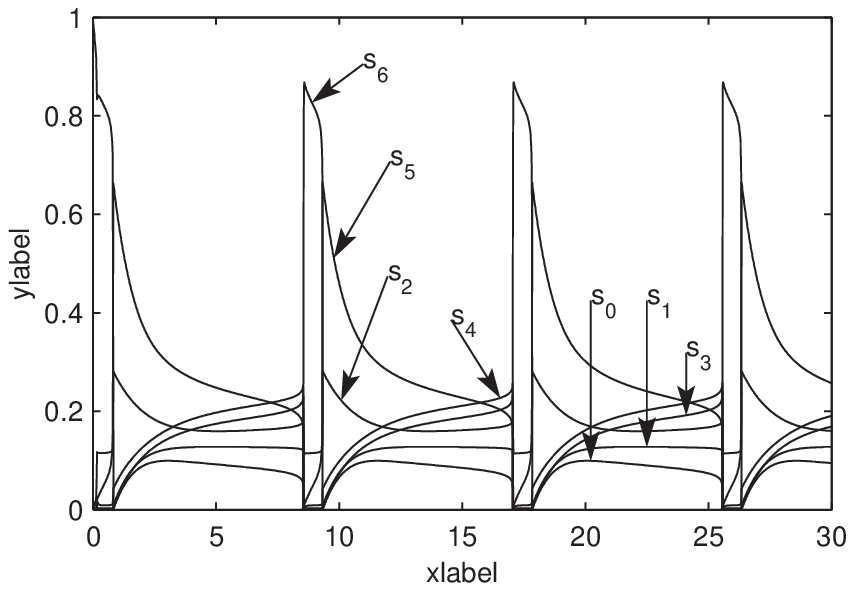}
	}
	\hspace{-7.5cm}\raisebox{4.5cm}{\textbf{b}}\hspace{7.5cm}
	}

\caption{\label{fig:avg_payoffs} 
\textbf{a} 
The time-averaged average payoff in the population as a function
of $P$, for $n = 6$, $T = 1.6$, and $\mu
= 10^{-4}$. Also shown is
the payoff in the limit of infinitesimal $\mu$ (thick line) and
the payoff $P$ for a population of pure all-defect (dashed line).
(Adapted from \cite{eriksson_lindgren05}.)
\textbf{b} 
The initial transient and two cycles of the attractor of the time evolution 
of the population for $P = 0.33$ (Adapted from \cite{eriksson_lindgren05}).
Within a cycle, strategy $s_6$ comes to dominate the population for a period of time. During the
part of the cycle where strategy $s_6$ is prominent, the system
approaches the population mixture which is evolutionarily stable
when mutations are infrequent. For the value of $P$ used here, it
corresponds to a mixture of strategies $s_2$ and $s_6$. For a
period of time, this mixture of strategies dominates the
population. After a while, however, $x_5$ starts to grow at the
expense of $x_6$, and after a while there is another sharp
transition where $x_5$ and $x_2$ grow and $x_6$ decay to very low
levels.
}
\end{figure}

In \cite{eriksson_lindgren05} we introduced a simple model for incorporating mutations as an explicit part of the evolutionary dynamics. The population is subject to selection as in
(\ref{eq:replicator_dynamics_simple}). In addition, a number $\mathcal{M}_{l \rightarrow j}$ of individuals per generation switch from strategy $s_l$ to strategy $s_j$ due to mutations. The mutations are assumed to be generated by uncorrelated stochastic events, e.g. by a Poisson process, in the process of reproduction.
The evolutionary dynamics then takes the form
\beq\label{eq:replicator_dynamics_stoch}
   x'_i &=& x_i + \delta \left( \frac{f_i}{\bar{f}} - 1 \right) x_i + \frac{1}{N} \sum_{j=0}^n ( \mathcal{M}_{j \rightarrow i} - \mathcal{M}_{i \rightarrow j} ).
\eeq
In a single generation, each player with strategy $s_i$ has an expected number $\delta f_i x_i / \bar{f}$ of offspring surviving to reproductive age. Mutations occur independently in the creation of each offspring, with a probability of $\mu \ll 1$ per offspring per generation, and the strategy of the mutated offspring is chosen randomly among the $n + 1$ strategies, with equal probability. Since the population size is assumed to be large, we approximate the number $\mathcal{M}_{i \rightarrow j}$ of mutated offspring with its expected value, $ \mathcal{M}_{i \rightarrow j} \approx \mu \delta N f_i x_i/[(n+1) \bar{f}]$. Inserting this approximation into (\ref{eq:replicator_dynamics_stoch}), we obtain the following expression for the evolutionary dynamics:
\beq\label{eq:replicator_dynamics}
   x'_i &=& x_i + \delta \left( \frac{f_i}{\bar{f}} - 1 \right) x_i + 
   		  \delta\,\mu \left( \frac{1}{n+1} - \frac{f_i}{\bar{f}}\,x_i \right).
\eeq
For some values of the payoff parameters, the strength of selection is so small compared to the flow mutants that the population does not converge to the evolutionary stable composition of strategies in the population. This is illustrated in Fig.~\ref{fig:avg_payoffs}a, where we show the asymptotic time-averaged average fitness in the population, for case of infrequent mutations \cite{molander92} and for (\ref{eq:replicator_dynamics}). For some values of $P$ the difference is clear; in this case, the dynamics either converge to a limit cycle (illustrated in Fig.~\ref{fig:avg_payoffs}b), or to a fixed point where three or more strategies coexist at significant levels.

In this article, we compare the results of the deterministic approximation (\ref{eq:replicator_dynamics}) to explicit simulation of (\ref{eq:replicator_dynamics_stoch}).
It is natural to assume that in the stochastic model, the mutations in individuals with strategy $s_i$ occur according to a Poisson process with rate $\lambda_i = \mu \delta N f_i x_i/[(n+1) \bar{f}]$. 
In each simulation step, the number of mutants of each strategy is drawn independently. For each mutation, the resulting strategy is drawn with uniform distribution and the fraction of the chosen strategy is incremented.

An additional source of stochastic fluctuations come from the selection of individuals from one generation to the next. The growth of the fraction $x_i$ of strategy $s_i$ in (\ref{eq:replicator_dynamics_simple}) may be viewed as a mean-field approximation of a more realistic stochastic growth process, in which the difference in fitness between two individuals is reflected in the probability distribution of their number of offspring. For simplicity, we model mutations as occurring separate from the selection process -- in our computer simulations, we first generate the next generation according to the selection model, and then impose the effect of mutations as described above.

The selection process is as follows: in each generation, an individual with strategy $s$ produces a number of offspring in relation to the fitness of strategy $s$. It is assumed that the total number of offspring is more than enough to replace the parent generation. The fraction of the offspring with strategy $s_i$ is then
\beq
	(1-\delta) x_i + \delta\, \frac{f_i}{\bar{f}}\, x_i ,
\eeq
which we recognize as the fraction of $s_i$ in the next generation in the case of an infinite population size. Since the environment has a finite carrying capacity, we sample $N$ individuals uniformly from the offspring of the parent generation. Hence, the joint distribution of the number $k_i$ of players with strategy $s_i$ in the next generation is multinomial:
\beq
  f(k_0, \dots, k_n)  &=& \frac{N!}{\prod_{i=0}^n k_i!} \, \prod_{i=0}^n \left[ (1-\delta) x_i + \delta\, \frac{f_i}{\bar{f}}\,  x_i  \right]^{k_i} ,
\eeq
provided the constraint $\sum_{i=0}^n k_i = N$ is fulfilled. In the limit of $N \rightarrow \infty$, the variance of $x'_i = k_i/N$ vanishes, and the dynamics becomes deterministic, as expected.

The selection process can be approximated, for large populations, with a diffusion process: to each fraction $x_i$ we add a normal distributed number $\xi_i$ with zero mean and a variance $\delta x_i (1 - x_i)/N$. We take the variables $\xi_i$ and $\xi_j$ to be uncorrelated, for all $i \ne j$. If a fraction $x_i$ becomes negative, it is set to zero. Finally, the fractions are normalised so that $\sum_{i=0}^n x_i = 1$. This approximation does not exhibit the correct correlations in the fluctuations of the $x_i$, but the magnitude of the fluctuations are approximately correct. Since the magnitude of the perturbations are small, however, correlations and higher-order cross-terms may be neglected.

\section{Results}
\label{sec:results}

\begin{figure}[tb]
\begin{tabular}{@{}l@{}l@{}}
\psfrag{t}{}
\psfrag{xi}[][t]{Fraction $x_i$}
\includegraphics[clip]{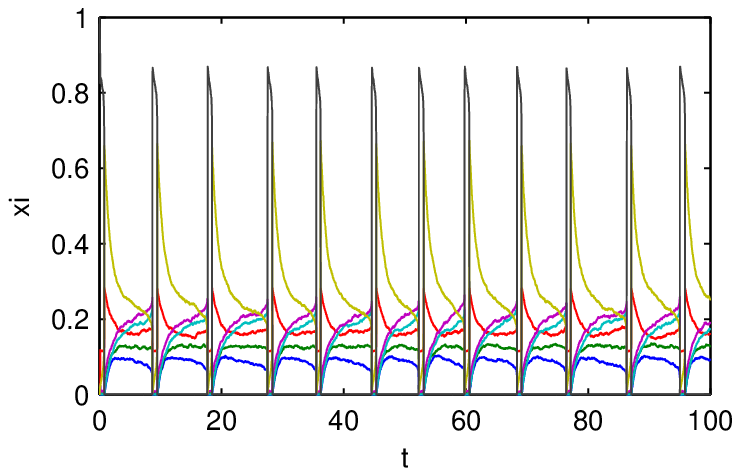} &
\hspace{-7.7cm}\raisebox{45mm}{\textbf{a}}\hspace{7.5cm}
\psfrag{xi}{}
\psfrag{t}{}
\includegraphics[clip]{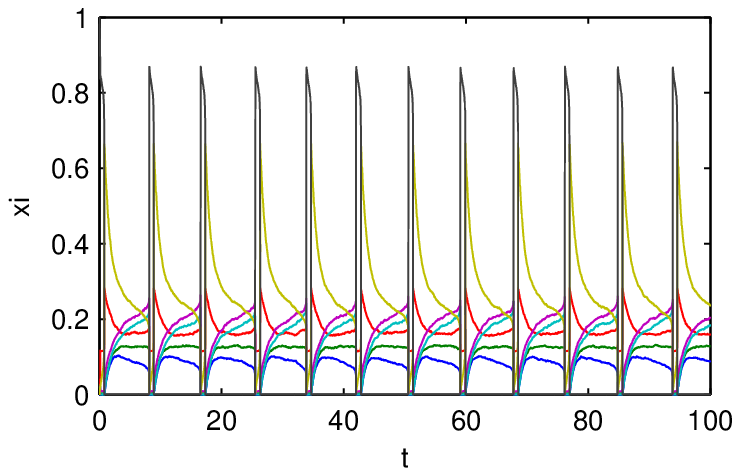} 
\hspace{-7.5cm}\raisebox{45mm}{\textbf{b}}\hspace{7.5cm}
\\[-2mm]
\psfrag{xi}[][t]{Fraction $x_i$}
\psfrag{t}[t][]{$\mu t$}
\includegraphics[clip]{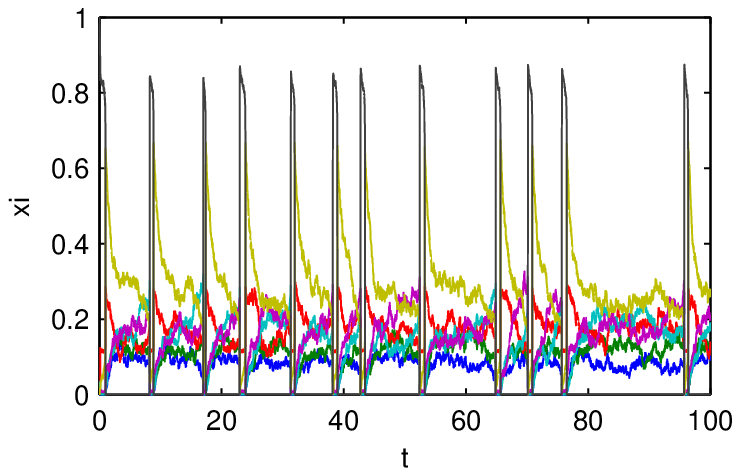} &
\hspace{-7.7cm}\raisebox{45mm}{\textbf{c}}\hspace{7.5cm}
\psfrag{xi}{}
\psfrag{t}[t][]{$\mu t$}
\includegraphics[clip]{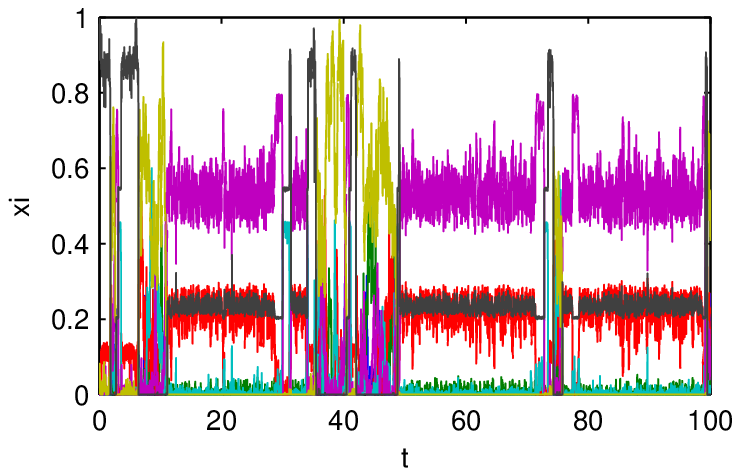} 
\hspace{-7.5cm}\raisebox{45mm}{\textbf{d}}\hspace{7.5cm}
\end{tabular}

\caption{\label{fig:pop_size_effect}
Illustration of the effect of stochastic mutations and a finite population size on the evolution of strategies. The parameters are: $T = 1.6$, $P = 0.33$, $\mu = 10^{-4}$ and the groups size $n = 6$. The lines are: $s_0$ (blue), $s_1$ (green), $s_2$ (red), $s_3$ (cyan), $s_4$ (magenta),  $s_5$ (yellow), and  $s_6$ (black).  In panel {\bf a} we show the time evolution of a population with stochastic mutations but ignoring the effect of finite population size in the selection (population size $N = 10^4$). In panels {\bf b}, {\bf c}, and {\bf d}, both explicit mutation and selection is present. The effective population size $N$ takes the value $10^8$, $10^6$, and $10^4$, respectively. 
}
\end{figure}

In Fig.~\ref{fig:pop_size_effect} we illustrate the effect of random mutations and the finite population size on the evolution of the frequencies of the strategies, for a specific choice of the parameters. The effect is similar for other parameters; for parameter values where the deterministic approximation (\ref{eq:replicator_dynamics}) converges to a fixed point (c.f. Fig.~\ref{fig:avg_payoffs}a), introducing a finite (large) population size does not change the qualitative properties of the long-term evolutionary dynamics, although the variance of the fraction of each strategy is increasing with decreasing population size.

We now focus on the fate of the limit cycles that are the most striking feature in the deviations of (\ref{eq:replicator_dynamics})  from the evolutionary dynamics in the limit of infinitesimal mutation rate. The simulations show that, in the limit of very large populations, the analytical approximation (\ref{eq:replicator_dynamics}) of how the flow of mutants alter the replicator dynamics is valid. As can be see from the smoothness of the curves in Fig.~\ref{fig:pop_size_effect}a, the main effect of the stochastic mutations compared to their deterministic counterpart is to add a little noise to the time evolution so that the curve fluctuates around the deterministic solution. Hence, unless the population size is small, the difference between the explicit mutations and the analytical approximation is negligible.

Fluctuations in the frequencies of strategies in the population due to random sampling in the selection play a very different role. Panels b--d in  Fig.~\ref{fig:pop_size_effect} illustrate the effect of these fluctuations for three values of the populations size. For very large populations ($N \sim 10^8$), the fluctuations do not alter the qualitative properties of the time evolution (panel b). However, for populations as large as $N=10^6$ individuals (panel c), we find that the fluctuations cause significant deviations from the  deterministic model, while some qualitative aspects are the same: the time evolution still exhibit cycles, characterised by short periods where the composition is similar to that of the infrequent-mutation solution of Molander \cite{molander92}, interspersed with long periods of a high degree of cooperation and with a slow drift of strategies. The period of these cycles now fluctuate. When the population size is decreased to $N=10^4$, the cycles are no longer stable. The population now switches between irregular cyclic behaviour and  more stable configurations in which three or more strategies co-exist. Preliminary investigations indicate that these configuration may have a stable counterpart in the deterministic approximation; it is not yet clear, however, if this he case for all parameter values.

\section{ Discussion and conclusions}

The main purpose of this article (and also of the previous article \cite{eriksson_lindgren05}) is to better understand how more realistic models of the evolution of a population affects the evolution of the frequencies of strategies in the population. In the present article we have confirmed that -- when the population size is large and the  mutation rate is small -- the deterministic approximation (\ref{eq:replicator_dynamics}) provides an accurate description of the evolutionary dynamics. We have also shown that taking the fluctuations in the selection process into account may lead to different compositions of the strategies in the population than expected from the earlier analysis.

The fluctuations in the selection of individuals show a much stronger effect on the evolutionary dynamics than does the fluctuations from the mutation process. A large part of this difference can be understood from the magnitudes of these fluctuations, as characterized by their variances: 
in the mutation process, the variance contributed per time step is of order $[\delta \mu x_i /(n+1)]N^{-1}$, whereas in the selection process the variance is $\delta x_i(1 - x_i)N^{-1}$. When the mutation rate is small, and $x_i$ is not very close to one, $\mu/(n+1) \ll 1 - x_i$. For instance, when $\mu = 10^{-4}$, $N = 10^4$, and $n = 6$, as in Fig.~\ref{fig:pop_size_effect}a, the typical fluctuations due to mutations are of the same order as the fluctuations from the selection process in population with $N \sim 10^8$ individuals. 
This explains the good qualitative agreement between panels a and b in Fig.~\ref{fig:pop_size_effect}.

It remains to explain what causes the evolutionary dynamics to switch from the limit cycle to coexistence of three strategies in Fig.~\ref{fig:pop_size_effect}d. A hypothesis is that, apart from the limit cycle,  there are stable fixed points with a small basin of attraction when the population evolves under (\ref{eq:replicator_dynamics}); the fluctuations in the selection process may then cause a transition from one mode to the other and back. In this case we expect that the transitions between these modes occur, approximately, according to a Poisson process. The number of such fixed points, and to which extent the proposed mechanism can explain the observed dynamics, remains to be investigated.

\section*{Acknowledgement}

The authors thank one of the reviewers for comments on the manuscript.

\bibliographystyle{unsrt}

\begin{thebibliography}{10}

\bibitem{hardin68}
G.~Hardin.
\newblock The tragedy of the commons.
\newblock {\em Science}, 162:1243--1248, 1968.

\bibitem{maynard-smith82}
J.~Maynard~Smith.
\newblock {\em Evolution and the Theory of Games.}
\newblock Cambridge University Press, Cambridge, 1982.

\bibitem{sugden86}
R.~Sugden.
\newblock {\em The Economics of Rights, Co-operation and Welfare}.
\newblock Basil Blackwell, Oxford, 1986.

\bibitem{flood58}
M.~M. Flood.
\newblock Some experimental games.
\newblock {\em Management Science}, 5(1):5--26, 1958.
\newblock First published in Report RM-789-1, The Rand Corporation, Santa
  Monica, CA, 1952.

\bibitem{rapoport_chammah65}
A.~Rapoport and A.~M. Chammah.
\newblock {\em Prisoner's Dilemma.}
\newblock University of Michigan Press, Ann Arbor, 1965.

\bibitem{axelrod_hamilton81}
R.~Axelrod and D.~H. Hamilton.
\newblock The evolution of cooperation.
\newblock {\em Science}, 211:1390 -- 1396, 1981.

\bibitem{molander85}
P.~Molander.
\newblock The optimal level of generosity in a selfish, uncertain environment.
\newblock {\em J. Conflict Resolution}, 29:611 -- 618, 1985.

\bibitem{axelrod87}
R.~Axelrod.
\newblock The evolution of strategies in the iterated prisoner's dilemma.
\newblock In Davis L, editor, {\em Genetic Algorithms and Simulated Annealing},
  pages 32 -- 41. Morgan Kaufmann, Los Altos, CA, 1987.

\bibitem{miller96}
J.~H. Miller.
\newblock The coevolution of automata in the repeated {Prisoner's Dilemma}.
\newblock {\em J. Econ. Behav. Organ.}, 29(1):87--112, 1996.
\newblock Appeared 1989 as a Santa Fe Institute working paper.

\bibitem{lindgren92}
K.~Lindgren.
\newblock Evolutionary phenomena in simple dynamics.
\newblock In {Langton, C. G. \emph{et al}}, editor, {\em Artificial Life II},
  pages 295 -- 311. Addison-Wesley, Redwood City, CA, 1992.

\bibitem{nowak_sigmund92}
M.~A. Nowak and K.~Sigmund.
\newblock Tit for tat in heterogenous populations.
\newblock {\em Nature}, 359:250--253, 1992.

\bibitem{matsuo85}
K.~Matsuo.
\newblock Ecological characteristics of strategic groups in 'dilemmatic world'.
\newblock In {\em IEEE International Conference on Systems and Cybernetics},
  pages 1071 -- 1075, 1985.

\bibitem{adachi_matsuo91}
N.~Adatchi and K.~Matsuo.
\newblock Ecological dynamics under different selection rules in distributed
  and iterated prisoner's dilemma games.
\newblock In {\em Parallel Problem Solving From Nature}, volume 496 of {\em
  Lecture Notes in Computer Science}, pages 388--394. Springer, Berlin, 1991.

\bibitem{adachi_matsuo92}
N.~Adatchi and K.~Matsuo.
\newblock Ecological dynamics of strategic species in game world.
\newblock {\em Fujitsu Sci. Tech. J.}, 195:543--558, 1992.

\bibitem{albin92}
P.~Albin.
\newblock Approximations of cooperative equilibria in multi-person {Prisoner's
  Dilemma} played by cellular automata.
\newblock {\em Math. Soc. Sci.}, 24:293--319, 1992.

\bibitem{hauert_schuster97}
C.~Hauert and H.~G. Schuster.
\newblock Effects of increasing the number of players and memory size in the
  iterated {Prisoner's Dilemma}, a numerical approach.
\newblock {\em Proc. R. Soc. B London}, 264:513--519, 1997.

\bibitem{akimov_soutchanski97}
V.~Akimov and M.~Soutchanski.
\newblock Automata simulation of $n$-person social dilemma games.
\newblock {\em J. Conflict Resolution}, 38:138--148, 1997.

\bibitem{matsushima_ikegami98}
M.~Matsushima and T.~Ikegami.
\newblock Evolution of strategies in the three-person iterated prisoner's
  dilemma game.
\newblock {\em J. Theor. Biol.}, 195:53--67, 1998.

\bibitem{lindgren_johansson01}
K.~Lindgren and J.~Johansson.
\newblock Coevolution of strategies in $n$-person {Prisoner's Dilemma}.
\newblock In J.~Crutchfield and P.~Schuster, editors, {\em Evolutionary
  Dynamics - Exploring the Interplay of Selection, Neutrality, Accident, and
  Function}. Addison-Wesley, 2001.

\bibitem{eriksson_lindgren05}
A.~Eriksson and K.~Lindgren.
\newblock Cooperation driven by mutations in multi-person prisoner's dilemma.
\newblock {\em J. Theor. Biol.}, 232(3):399--409, 2005.

\bibitem{boyd_richerson88}
R.~Boyd and P.~Richerson.
\newblock The evolution of reciprocity in sizable groups.
\newblock {\em J. Theor. Biol.}, 132:337--356, 1988.

\bibitem{molander92}
P.~Molander.
\newblock The prevalence of free riding.
\newblock {\em J. Conflict Resolution}, 36:756--771, 1992.

\bibitem{schelling78}
T.~C. Schelling.
\newblock {\em Micromotives and Macrobehaviour}.
\newblock Norton, New York, 1978.

\end{thebibliography}

\end{document}